\documentstyle[aps,preprint,tighten,epsf]{revtex}

\begin{document}

\preprint{}
\draft

\title{
An Exactly Solvable Model of $N$ Coupled Luttinger Chains}
\author{Lorenz Bartosch and Peter Kopietz}
\address{
Institut f\"{u}r Theoretische Physik der Universit\"{a}t G\"{o}ttingen,\\
Bunsenstr.9, D-37073 G\"{o}ttingen, Germany}
\date{Fabruary 13, 1997}
\maketitle
\begin{abstract}
We calculate the exact Green function of a special model of $N$
coupled Luttinger chains with arbitrary interchain hopping $t_{\perp}$.
The model is exactly solvable via bosonization if the interchain
interaction does not fall off in the direction perpendicular to the
chains. For any finite $N$ we find Luttinger liquid behavior and explicitly
calculate the anomalous dimension $\gamma^{(N)}$. However, the
Luttinger liquid state does not preclude coherent interchain
hopping. We also show that $\gamma^{(N)} \to 0$ for $N \to \infty$, so
that in the limit of infinitely many chains we obtain a Fermi liquid.
\end{abstract}
\narrowtext

The normal metallic state of interacting
electrons in one spatial dimension is called a Luttinger
liquid, and has fundamentally different properties
than the Fermi liquid state in higher-dimensional Fermi systems.
In particular, the single-particle Green function
of a Luttinger liquid does not have
a quasi-particle peak,
but exhibits spin-charge separation and
anomalous scaling properties.
In order to calculate the physical properties of Luttinger liquids,
non-perturbative methods are necessary, 
such as the bosonization method \cite{bosonisation}.
Recently the crossover from Luttinger liquid to Fermi liquid
behavior as function of some suitable parameter has received much
attention \cite{chains}, partially because of Anderson's and
coworkers' suggestion 
that higher-dimensional Luttinger liquid behavior
might be the key to understand high-temperature superconductivity.
An experimentally important class of systems 
exhibiting the above mentioned crossover
are quasi-one-dimensional chain-like 
metals. In the normal metallic state (which is realized at sufficiently
high temperatures), these systems can be described as one-dimensional
Luttinger liquids
that are coupled by small interchain hopping $t_{\bot}$ and by the
three dimensional interaction.
While for $t_{\bot} = 0$ the number of particles on each chain is 
conserved and
the problem can be studied in a straightforward way via bosonization,
the case of $t_{\bot} \neq 0$ is more difficult and has recently been
studied by means of a variety of non-perturbative methods.
Most authors have focused on the crossover as a function of
the dimensionless parameter $ t_{\bot} / E_F$ (where $E_F$ is the Fermi 
energy),
assuming that this parameter is small.
In this work we shall show that for some special type of
interactions (see below) it is possible to obtain
an exact solution for the Green function of an {\it{arbitrary
number of coupled Luttinger chains with finite $t_{\bot}$.}}
The interaction is characterized by the fact that it does not fall off
in the transverse direction.
Although such an interaction is unphysical, 
our model is useful for testing the various approximations employed
in the literature.

\begin{figure}[htb]
  \begin{center}
    \epsfysize6cm
    \leavevmode
    \epsfbox{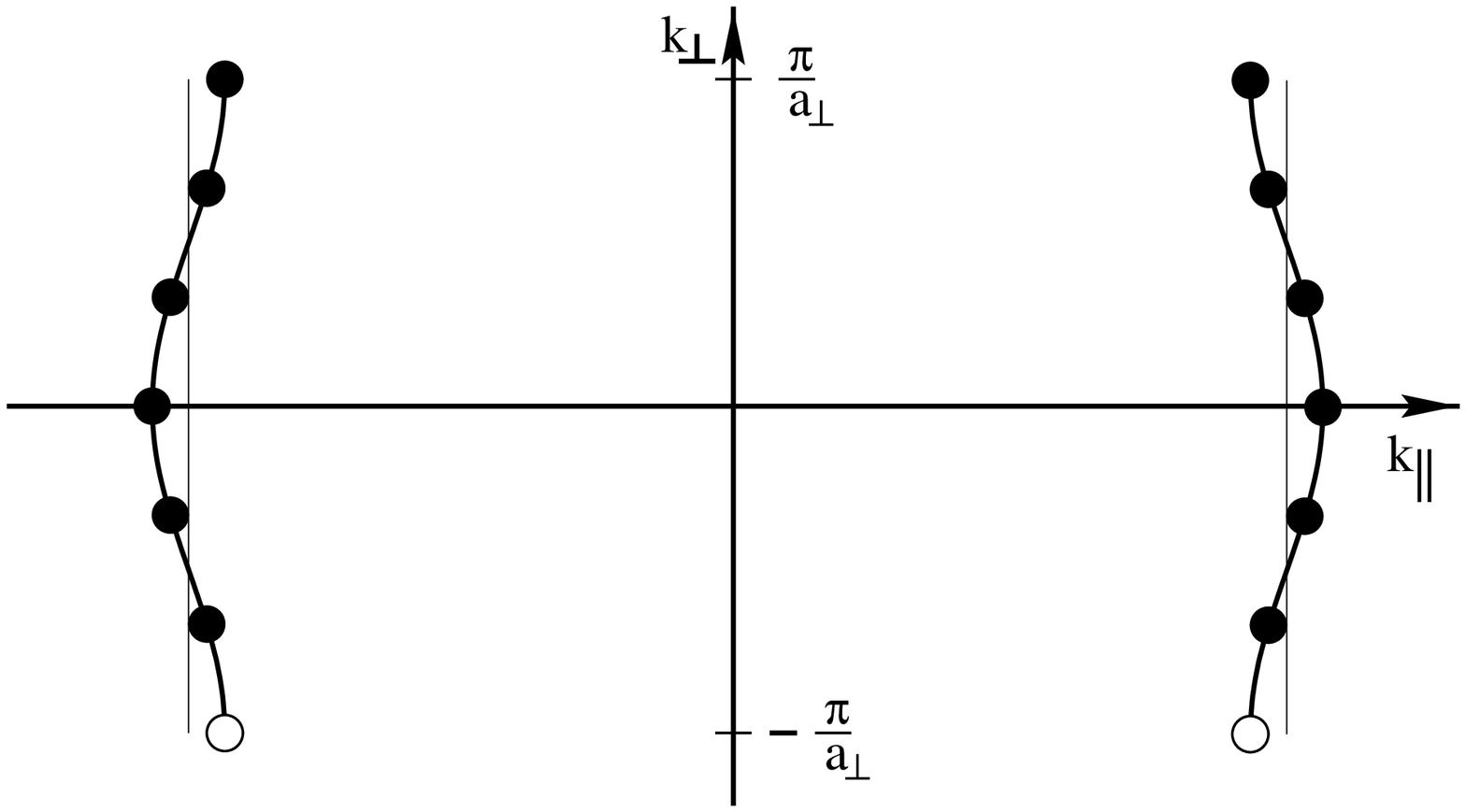}
    \caption{Fermi-surface for the model with $N = 6$ chains and
    periodic boundary conditions.}
    \label{fermisurface}
  \end{center}
\end{figure}

Let us start with a two-dimensional system of interacting fermions
moving on an array of $N$ weakly coupled spin $1/2$ chains in the
$x$-direction. Each chain can be described by the Tomonaga-Luttinger
model with spin \cite{tomonaga}.
Denoting by $k_{\|}$ the momentum along the chains and 
linearizing the energy dispersion in the chain direction,
the kinetic energy part of the Hamiltonian can be written as
 \begin{equation}
  H_{\rm kin} = \sum_{i,j}\sum_{\alpha,\sigma} \sum_{k_{\parallel}}
  \left[\delta_{i,j} \epsilon^{\alpha}(k_{\parallel}) -
  t_{ij}\right] \left(\hat c_{i \sigma}^{\alpha
  \dag}(k_{\parallel})\hat c_{j \sigma}^{\alpha}(k_{\parallel}) - 
  \langle \hat c_{i \sigma}^{\alpha
  \dag}(k_{\parallel})\hat c_{j \sigma}^{\alpha}(k_{\parallel})
  \rangle_{0}\right)
  \label{eq:Hkindef}
  \; \; \; ,
  \end{equation}
where $\hat c_{i \sigma}^{\alpha \dag}(k_{\parallel})$ and $\hat c_{i
  \sigma}^{\alpha}(k_{\parallel})$ are the creation and annihilation
  operators for the right and left moving fermions $(\alpha = +,-)$ on
  chain $i=[-\frac{N}{2}+1],\ldots,[\frac{N}{2}]$ with spin
  $\sigma = \uparrow,\downarrow$ and momentum
  $k_{\parallel}$ in the chain direction. 
The linearized energy dispersions are
$\epsilon^{\pm}(k_{\parallel}) =
v_F(\pm k - k_F)$, where $v_F$ is the Fermi velocity and
$k_F$ is the Fermi momentum associated in the absence of interchain
hopping.
For simplicity we assume only 
nearest neighbor hopping and impose periodic boundary conditions
in the transverse direction. In this case the
interchain hopping is for $N >2$ of the form
$t_{ij} = t_{i-j,0} = 
\delta_{|i-j|,1} t_{\perp}/2$. In the special case of just two chains periodic
boundary conditions are automatically satisfied and we set $t_{ij} =
t_{\perp} \delta_{|i-j|,1}$. 
Because of the discrete translational invariance in the
transverse direction (the $y$-direction) we can
completely diagonalize $H_{\rm kin}$ 
via a discrete Fourier transformation in the
transverse direction. Denoting by
$k_{\perp}$ the
corresponding transverse momentum and defining
 $\hat c_{\sigma}^{\alpha}({\bf k}) =
    N^{-1/2} \sum_{i} e^{-ik_{\perp} y_i}
    \hat c_{i\sigma}^{\alpha}(k_{\parallel})$, we have
 \begin{equation}
  H_{\rm kin} = \sum_{\alpha,\sigma} \sum_{{\bf k}} \epsilon^{\alpha}({\bf
  k}) \left[\hat c_{\sigma}^{\alpha \dag}({\bf k})\hat
  c_{\sigma}^{\alpha}({\bf k}) - \langle \hat c_{\sigma}^{\alpha
  \dag}({\bf k})\hat
  c_{\sigma}^{\alpha}({\bf k}) \rangle_0\right]
  \; \; \; ,
  \label{eq:Hkin2}
  \end{equation}
where
${\bf k} = (k_{\parallel},k_{\perp})$, and
the ${\bf k}$-dependent energy dispersion is given by $
\epsilon^{\alpha}({\bf k}) = \epsilon^{\alpha}(k_{\parallel}) - 
t ( k_{\perp} )$, with $t ( k_{\perp} ) =
  t_{\perp}\cos(k_{\perp} a_{\perp})$. Here $a_{\bot}$ is the distance
between the chains.  The corresponding Fermi surface
is shown in Fig.\ref{fermisurface}.

The total Hamiltonian is  given by $H = H_{\rm kin} + H_{\rm int}$, where
the interaction part is in general of the form
 \begin{equation}
  H_{\rm int} = \frac{1}{2}\frac{1}{2L} \sum_{i,j}\sum_{\alpha,\alpha'}
  \sum_{\sigma,\sigma'}\sum_{q_{\parallel}} f_{ij\sigma \sigma'}^{\alpha
  \alpha'} (q_{\parallel})
  :\hat\rho_{i\sigma}^{\alpha\dag}(q_{\parallel})
  \hat\rho_{j\sigma'}^{\alpha'}(q_{\parallel}):\quad. 
  \label{eq:Hint}
  \end{equation}
Here, one factor of $1/2$ has been introduced for convenience due to the spin, and
the Fourier components of the density operators associated with chain
$i$ are given by
 \begin{equation}
 \hat \rho^\alpha_{i\sigma}(q) =
 \sum_k \left(\hat c^{\alpha \dag}_{i\sigma}(k)\hat
 c^{\alpha}_{i\sigma}(k+q) -  \left< \hat c^{\alpha \dag}_{i\sigma}(k)\hat
 c^{\alpha}_{i\sigma}(k+q)
 \right>_{0} \right)
 \; \; \; .
 \end{equation}
Assuming spin and inversion symmetry, we
thus assume that the interaction parameters satisfy
$f^{++}_{\sigma\sigma'} = f^{--}_{\sigma\sigma'}$, 
$f^{+-}_{\sigma\sigma'} = f^{-+}_{\sigma\sigma'}$, 
$f_{\parallel}^{\alpha\alpha'} \equiv
  f_{\uparrow\uparrow}^{\alpha\alpha'} =
  f_{\downarrow\downarrow}^{\alpha\alpha'}$,
  and $f_{\perp}^{\alpha\alpha'} \equiv
  f_{\uparrow\downarrow}^{\alpha\alpha'} =
  f_{\downarrow\uparrow}^{\alpha\alpha'}$. 
Performing the complete Fourier transformation of
$H_{\rm int}$ by defining
 \begin{eqnarray}
  \hat\rho_{\sigma}^{\alpha}({\bf q})
  & = &
  \frac{1}{\sqrt{N}}\sum_{{\bf k}} \Big(\hat
  c_{\sigma}^{\alpha\dag}({\bf k}) \hat
  c_{\sigma}^{\alpha}({\bf k}+{\bf q})-\langle  \hat
  c_{\sigma}^{\alpha\dag}({\bf k}) \hat
  c_{\sigma}^{\alpha}({\bf k}+{\bf q}) \rangle_0\Big)
  \; \; \; ,
  \label{eq:rhosigmaq}
  \\
  f_{\sigma \sigma'}^{\alpha
  \alpha'}({\bf q}) & =  &
 \sum_{i} e^{-iq_{\perp}y_{i}} 
  f_{i0\sigma\sigma'}^{\alpha\alpha'}(q_{\parallel})
  \label{eq:fqFT}
  \; \; \; ,
  \end{eqnarray}
and going over to a charge/spin basis,
 \begin{eqnarray}
 \hat \rho_{c/s}^{\alpha}({\bf q}) &=& \frac{1}{\sqrt{2}}\left[\hat
 \rho_{\uparrow}^{\alpha}({\bf q}) \pm \hat
 \rho_{\downarrow}^{\alpha}({\bf q}) \right]\quad,\\  
 f_{c / s}^{\alpha\alpha'}({\bf q}) &=&
 \frac{1}{2} \left[f_{\parallel}^{\alpha\alpha'}({\bf q}) \pm
 f_{\perp}^{\alpha\alpha'}({\bf q}) \right]
 \; \; \; ,
 \end{eqnarray}
we obtain
\begin{equation}
  H_{\rm int} = \frac{1}{2L} \sum_{\alpha,\alpha'}
  \sum_{\nu=c,s}\sum_{{\bf q}}
  f_{\nu}^{\alpha\alpha'}({\bf q}) :\hat\rho_{\nu}^{\alpha
  \dag} ({\bf q}) \hat\rho_{\nu}^{\alpha'} ({\bf q}):\quad.
\end{equation}

\begin{figure}[htb]
  \begin{center}
    \epsfysize6cm
    \leavevmode
    \epsfbox{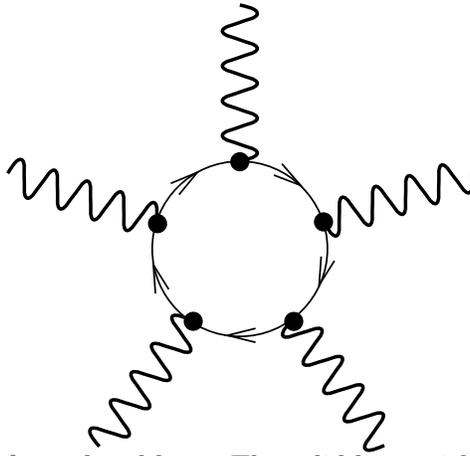}
    \caption{Feynman diagram for a closed loop. The solid lines with
    an arrow represent free Green functions and the wiggled lines
    represent the auxiliary field $\phi$. The closed loop theorem
    states that the contributions from all diagrams with more than two
    auxiliary fields in a closed loop vanish.}
    \label{closedloop}
  \end{center}
\end{figure}

We would like to calculate the single-particle Green function of the
model defined above. In particular, we are interested in the
fate of the Luttinger liquid state as function of $t_{\bot}$ and $N$.
Note that for $t_{\bot} =0$ the above model is exactly solvable
by means of bosonization, because in this case the particle
number on each chain is conserved \cite{schulz}.
The case of finite $t_{\bot}$ is very difficult to handle, and there
exists no complete agreement in the literature about the nature
of the ground state.
In this work we would like to point out that there exists a special type
of interaction
$f^{\alpha\alpha'}_{\nu}({\bf q})$ where the above model 
is exactly solvable via bosonization {\it{for arbitrary
$t_{\bot}$ and $N$}}. This is easily seen
in the functional bosonization approach\cite{leechen,peter},
where  the interaction part of the effective action
corresponding to $H_{\rm int}$ is decoupled via a
dynamic Hubbard-Stratonovich field $\phi$.
In this approach the exact solubility of the Tomonaga-Luttinger model
manifests itself via the fact that in the perturbative expansion of
the effective action
for the $\phi$-field, which is obtained by
integration over the fermionic degrees of freedom in the
usual way, all non-Gaussian terms vanish identically (see Fig.\
\ref{closedloop}). 
This cancellation has first been noticed by
Dzyaloshinski\v{i} and Larkin \cite{dzyaloshinskii}, and was later discussed in detail
by T. Bohr \cite{bohr}, who formulated this calculation in terms
of a theorem which he called
{\it{closed loop theorem}}.
While this theorem is exact in the 
one-dimensional Tomonaga-Luttinger model, 
it remains approximately valid even in higher dimensions
if the interaction is dominated by forward scattering \cite{peter,metzner}.

The crucial observation of this work is that
for our model defined above 
the closed loop theorem 
is still exact for interactions of the type
$f_{\nu}^{\alpha\alpha'}({\bf q}) = 0$ for $q_{\perp} \not= 0$,
which in real space amounts to a potential which is independent of the
chain indices.
In this case
the auxiliary fields $\phi$
do not transfer any {\it{transverse}} momentum into closed fermion loops.
For a linearized energy dispersion along the chain direction,
the loops with more than two external fields
cancel then {\it{for exactly the same reason as in the
one-dimensional Tomonaga-Luttinger model.}}
After Fourier transformation this system is essentially
equivalent to
a system consisting of $N$ independent
Luttinger chains each with a different
Fermi-momentum and 
with only one Luttinger chain showing interaction.
We would like to emphasize that this model is exactly solvable
for arbitrary $N$ and $t_{\bot} $, although it is physically meaningful
only for $|t_{\bot}| \ll E_F$, because we have linearized the energy
dispersion along the chain direction.

The calculation of the imaginary-time Green function is now analogous to the
calculation of the Green function of the
Tomonaga-Luttinger model.
For simplicity we set the chemical
potential $\mu = 0$. In this case there is no need for a special
treatment of the particle mode \cite{lorenz}. 
Following a suggestion of Luther and Peschel
\cite{lutherpeschel}, we assume a potential which satisfies
$\gamma_{\nu}(q_{\parallel}) = \gamma_{\nu} e^{-r|q_{\parallel}|}$, where 
 \begin{eqnarray}
 \gamma_{\nu}(q_{\parallel}) & = &
 \frac{1}{2}\left[\frac{v^{+}_{\nu}(q_{\parallel},0)}{v_{\nu}(q_{\parallel},0)} 
- 1\right]
 \quad ,
 \label{eq:gammadef}
 \\
 v^{+}_{\nu}({\bf q}) & = & v_F + \frac{f^{++}_{\nu}({\bf q})}{2\pi}
 \quad ,
 \label{eq:vpl}
 \\
 v_{\nu}({\bf q}) & = &
 \sqrt{(v^{+}_{\nu}({\bf q}))^{2} -
 (\frac{f^{+-}_{\nu}({\bf q})}{2\pi})^{2} }
 \quad .
 \label{eq:vq}
 \end{eqnarray}
For $L \rightarrow \infty$  
we obtain for the real-space imaginary-time Green function
at finite temperatures
\begin{equation}
  {{\cal G}}_{\sigma}^{\alpha}(x,k_{\perp},\tau) =
  {{\cal G}}_{0}^{\alpha}(x,\tau) e^{i\alpha \frac{t(k_{\perp})}{v_F} x}
  \prod_{\nu=c,s}\Biggl[\exp\left(Q_{\nu}^{\alpha}(x,\tau)\right)\Biggr]^{\frac{1}{2N}}, 
\end{equation}
where
\begin{eqnarray}
  {\cal G}^{\alpha}_0(x,\tau) &=&  - \frac{\alpha}{2\pi i}\
  \frac{(\pi/\beta v_F)e^{i\alpha k_F x}}{\sinh[(\pi/\beta
  v_F)(x+i\alpha v_F \tau)]}\quad,  
  \\
  \exp\left(Q_{\nu}^{\alpha}(x,\tau)\right) 
  &=& 
  L_{\nu}(x,\tau)
  K_{\nu}(x,\tau) 
  \frac{x+i\alpha v_F \tau
  +i\alpha r}{x+i\alpha v_{\nu} \tau +i\alpha r}
  \nonumber
  \\
  & \times &
  \left[\frac{r^2}{(x+i v_{\nu} \tau +ir)
  (x-iv_{\nu} \tau
  -ir)}\right]^{\gamma_{\nu}} \quad ,  \\
  L_{\nu}(x,\tau) &=& \frac{\Gamma(a_{\nu} +
  iu_{\nu}^{\alpha}) \Gamma(a_{\nu} - iu_{\nu}^{\alpha})}{\Gamma(a_F +
  iu_F^{\alpha}) \Gamma(a_F - iu_F^{\alpha})}\quad,  \\
  K_{\nu}(x,\tau) &=& \left[\frac{\Gamma(a_{\nu} +
  iu_{\nu}^{+}) \Gamma(a_{\nu} - iu_{\nu}^{+})\Gamma(a_{\nu} +
  iu_{\nu}^{-}) \Gamma(a_{\nu} - iu_{\nu}^{-})}{\Gamma(a_{\nu}) \Gamma(a_{\nu})
  \Gamma(a_{\nu}) \Gamma(a_{\nu})} \right]^{\gamma_{\nu}}\quad,
\end{eqnarray}
and $a_i = 1+{r}{\beta v_i}, u_i^{\alpha} =
(x+i\alpha v_i \tau)/(\beta v_i) $. 
Note that at zero temperature both factors $L_{\nu}(x,\tau)$ and
$K_{\nu}(x,\tau)$ are equal to one. 
Because the only $k_{\perp}$-dependence is in the $e^{i\alpha
  \frac{t(k_{\perp})}{v_F} x}$-factor, this Green function is easily
transformed to the real-space basis in the transverse direction,
\begin{equation}
  {\cal G}_{j-l,\sigma}^{\alpha}(x,\tau) =
  {\cal G}_0^{\alpha}(x,\tau) 
  g_{j-l} ( x )
  \prod_{\nu=c,s}\Biggl[\exp\left(Q_{\nu}^{\alpha}(x,\tau)\right)\Biggr]^{\frac{1}{2N}}
  \label{eq:Gsol}
  \; \; \; ,
\end{equation}
with
 \begin{equation}
  g_{j-l} ( x )
  = \frac{1}{N}\sum_{k_{\perp}}
 e^{i k_{\perp}a_{\perp}(j-l) + i\alpha\frac{t_{\perp}}{v_{F}}x 
 \cos(k_{\perp}a_{\perp})}
 \label{eq:gijx}
 \; \; \; .
 \end{equation}
This is the exact expression for the Green function in real
space and imaginary time. The only $t_{\perp}$-dependence is in the
factor $g_{j-l} ( x )$.
In the 
special case of just two chains this factor simplifies for $j=l$ to
$\cos\left(\frac{t_{\perp}}{v_F} x\right)$ and for $j\not=l$ to
$(i\alpha)\sin\left(\frac{t_{\perp}}{v_F} x\right)$.

The above result has a number of interesting properties.
For any finite $N$ the Green function in
Eq.\ (\ref{eq:Gsol}) shows typical Luttinger liquid behavior. 
There are different possibilities of scaling the interaction as the
number of chains increases. Analogous to the Weiss-model one could
choose $f^{\alpha\alpha'}_{ij\nu}(q_{\parallel}) =
\frac{f^{\alpha\alpha'}_{\nu}(q_{\parallel})}{N}$, where
$f_{\nu}^{\alpha\alpha'}(q_{\parallel})$ is independent of 
$N$. In this case 
$f_{\nu}^{\alpha\alpha'}({\bf q}) = \delta_{q_{\perp},0} 
f_{\nu}^{\alpha\alpha'}(q_{\parallel})$ is independent of $N$, and for
$N \to \infty$ the
anomalous dimension trivially scales to zero. Then the Green function of
the interacting model reduces to the Green function of the
noninteracting model. We will now show that this is even true for a
model for which the interaction
between two given chains does not depend on the total number of
chains $N$, i.e.\ $f^{\alpha\alpha'}_{ij\nu}(q_{\parallel}) =
f^{\alpha\alpha'}_{\nu}(q_{\parallel})$. In this case we have 
$f_{\nu}^{\alpha\alpha'}({\bf q}) = \delta_{q_{\perp},0} N
f_{\nu}^{\alpha\alpha'}(q_{\parallel})$.
The anomalous dimensions of the charge and spin channel can then 
be read off from Eq.\ (\ref{eq:gammadef}--\ref{eq:vq}) and Eq.\ (\ref{eq:Gsol}),
\begin{equation}
  \gamma_{\nu}^{(N)} = \frac{1}{2N}\left[\frac{v_F + N
  \frac{f_{\nu}^{++}(0)}{2\pi}}{\sqrt{\left({v_F + N
  \frac{f_{\nu}^{++}(0)}{2\pi}}\right)^2 -
  \left(N\frac{f_{\nu}^{+-}(0)}{2\pi}\right)^2 }} - 1 \right]\quad.
\end{equation}
For $N \rightarrow \infty$ 
the anomalous dimensions still vanish.
For $f^{++}_{\nu} = f^{+-}_{\nu}$ we have
$\gamma_{\nu}^{(N)} \propto N^{-1/2}$, while
$\gamma_{\nu}^{(N)} \propto N^{-1}$
for $f^{++}_{\nu} > f^{+-}_{\nu}$. In fact, it is easy to show that
in both cases 
 \begin{equation}
 \lim_{N\to\infty}
 [\exp\left(Q_{\nu}^{\alpha }(x,\tau)\right)]^{\frac{1}{N}} = 1
 \; \; \; ,
 \label{DWvanish}
 \end{equation}
so that in this limit the electron-electron interaction does not
affect the  Green function at all. Then Eq.\ (\ref{eq:Gsol})
reduces to the exact non-interacting
result with hopping,
\begin{equation}
   \lim_{N \rightarrow \infty} 
   \left[ {\cal G}_{j-l,\sigma}^{\alpha}(x,\tau)\right] 
   = {\cal G}_0^{\alpha}(x,\tau) 
   (i\alpha)^{|j-l|} J_{|j-l|}\left(\frac{t_{\perp}}{v_F}x\right) \quad,
\end{equation}
where $J_{p}\left(a\right)$ is the Bessel function of first kind and
order $p$. Obviously, we are not expanding in powers of $t_{\bot}$.
Thus, while for any finite $N$ the Green function shows Luttinger liquid
behavior, for $N \rightarrow \infty$ it reduces to the  Green function
of a trivial Fermi liquid.

To understand this
at the first sight rather surprising result,
consider the leading self-energy correction 
$\Sigma_{\nu}^{\alpha} ( {\bf{k}} , i \tilde{\omega}_n )$
in an expansion in powers of the RPA (random-phase approximation) interaction.
Assuming for simplicity that
$f_{\nu}^{\alpha\alpha^{\prime}}$ is independent of $\alpha$ and
$\alpha^{\prime}$ and omitting these indices, we have
\begin{equation}
  \Sigma_{\nu}^{\alpha} ({\bf k},i\tilde\omega_{n}) =
  -\frac{1}{\beta NL} \sum_{{\bf q},\omega_m} \frac{
  f_{\nu}({\bf q})  }{\epsilon ( {\bf{q}} , i \omega_m ) }
  G_{0}^{\alpha}({\bf k} - {\bf q},i\tilde\omega_{n}-i\omega_m)\quad,
  \label{eq:sigma}
\end{equation}
where $\tilde\omega_{n}$ is a fermionic and $\omega_m$ a bosonic
Matsubara frequency, 
  $G_{0}^{\alpha}({\bf k},i\tilde\omega_{n}) =
  [i\tilde\omega_{n} - \epsilon^{\alpha}({{\bf k}})]^{-1}$, and the
RPA dielectric function is given by
\begin{equation}
  \epsilon ( {\bf q},i\omega_m) = 
  1 + \frac{f_{\nu} ( {\bf{q}} )}{\pi v_F }  \frac{  v_F^{2} q_{\|}^2}{ 
  \omega_m^2 + v_F^2 q_{\parallel}^2}
  \quad .
  \label{eq:eRPA}
\end{equation}
Note that the closed loop theorem guarantees that all
corrections to 
the dielectric function beyond the RPA cancel.
Our model corresponds to
an interaction $f_{\nu} ( {\bf{q}} ) = \delta_{q_{\bot} , 0 } N f_{\nu} ( q_{\|} )$.
Hence only the
$q_{\parallel}$-summation in Eq.\ (\ref{eq:sigma}) 
survives, and the prefactor of $1/N$ 
is canceled by the factor $N$ in the bare interaction.
The above self-energy reduces then to the
self-energy of the Tomonaga-Luttinger model, {\it{except that the
effective dielectric function diverges in the limit $N \rightarrow \infty$.}}
Thus, while for any finite $N$ the perturbative self-energy of our model
exhibits the same type of singularities as the Tomonaga-Luttinger model,
in the limit $N \rightarrow \infty$
the effective interaction is completely screened, so that the lowest order 
self-energy vanishes. Our exact solution shows that this is
true to all orders in perturbation theory.

The strong screening of the interaction is clearly a consequence of the fact that
our model has a super-long-range bare interaction that does not
fall off in the transverse direction.
Although interactions of this type are unphysical, 
we have learned
two conceptually important points from our calculation.
First of all, a system consisting of a finite number of chains can
exhibit qualitatively different behavior from a system of infinitely
many chains. In particular, conclusions drawn from two coupled
Luttinger chains are in general not applicable to the physical more
relevant case of infinitely many chains.
In our model, we obtain Luttinger liquid behavior for any finite $N$,
but a trivial Fermi liquid for $N \rightarrow \infty$.
As shown above, for large $N$, the anomalous dimension vanishes as
$N^{-1/2}$ for $f^{++} = f^{+-}$, or as $N^{-1}$ for $f^{++} > f^{+-}$.
In numerical investigations
this scaling of the anomalous dimension with the number $N$ of coupled chains
could be a useful guide for the extrapolation to the infinite chain limit.
Finally, we would like to point out 
that two exactly solvable two-chain models that 
have independently been proposed by
Shannon, Li and d'Ambrumenil\cite{shannon} can be obtained as
special cases of our more general model with two chains ($N = 2$) by
choosing special types of interactions. 
As first noticed by Shannon {\it{et al.}}, these models
have the remarkable property of exhibiting Luttinger liquid behavior together
with coherent intrachain hopping. 
This is also true for our more general model of $N$ coupled 
Luttinger chains with spin,
as is evident from the factorized form
of the Green function given in Eq.\ (\ref{eq:Gsol}).
Thus, it is in general {\it{not correct}} that Luttinger liquid behavior
in an array of $N$ coupled chains precludes coherent single-particle
hopping
. However, this might be a special feature of our 
model, which has the for a large number of chains unrealistic property
that the interaction does not fall off in the transverse
direction. 
As shown here, the two-chain models proposed by Shannon {\it et al}
belong to the general class of models of coupled chains that are
characterized 
by an interaction which does not transfer any transverse
momentum. These models are equivalent to a system of $N$ {\it
  independent} one-dimensional chains. We believe that the coexistence
of coherent interchain hopping and Luttinger-liquid behavior is a
special feature of these models and is not relevant to physically more
realistic models where the interaction has a finite range in the
transverse direction \cite{Clarke94}.

We would like to thank Kurt Sch\"{o}nhammer for helpful discussions.

\end{document}